# Near-Unity Excitation and Radiative Efficiencies in Electroluminescence Without External Carrier Injection

Rui Li[1], Xinrui Li[1], Jiachen Xie[1], Pingjun Xu[1], Wen Li[1], Congcong Liu[1], Zheng Ge[1], Longjia Wu[2], Xingtong Chen[1], Zheming Liu[1], Song Chen[1,3]*

**Affiliations:**

[1]State Key Laboratory of Bioinspired Interfacial Materials Science, Chemical Engineering and Materials Science, Soochow University, Suzhou, Jiangsu 215123, China

[2]TCL Corporate Research, 1001 Zhongshan Park Road, Nanshan District, Shenzhen 518067 Guangdong, China.

[3]Jiangsu Key Laboratory of Advanced Negative Carbon Technologies, Soochow University, Suzhou, Jiangsu 215123, China.

*Song Chen. Email: songchen@suda.edu.cn

**Abstract:** Electroluminescence occurring without external charge injection is typically characterized by weak emission and excessive driving voltage, due to low excitation and radiative recombination efficiencies. Here, we demonstrate non-injecting electroluminescence (NI-EL) that challenges this conventional perception. To achieve this, we introduce an operational paradigm that leverages remote, state-abundant charge reservoirs, which elevates the excitation efficiency close to unity—a greater-than-20-fold improvement over the benchmark. This strategy is augmented by quantum dots (QDs) with nonmonotonically graded shells, raising the high-field radiative efficiency by approximately 7-8-fold. The resultant RGB NI-EL devices uniformly exhibit bright and efficient pulsed emissions, with key metrics including: a turn-on threshold of 3.7 $V_{rms}$ for red; a luminance of 291,628 nits and a power efficiency of 302.6 lm/W for green, with light-outcoupling enhancement; and, for blue, the superior emitter stability of the first QD-based blue NI-EL over its light-emitting diode counterpart. The minimal dielectric loss, sub-100 ns response time, and external EL efficiency of up to 45.4% further reinforce the mechanism–performance causality. These results overcome the inherent mechanistic limitations of NI-EL and establish performance that rivals or surpasses injection-type EL, including AC- or DC-driven variants, positioning NI-EL as a promising platform for high-performance pulsed light sources.

## Introduction

Since Gudden and Pohl's initial report in 1920, the development of electroluminescence (EL) technology has undergone multiple phases involving diverse excitation mechanisms and operational paradigms[1-3]. By the early 1990s, non-injecting (or capacitive coupled) electroluminescence (NI-EL)—which operates via electric field excitation without external charge injection—had become predominant in both research and commercial markets, attributable primarily to its superior reliability, lower dielectric loss, and simpler circuit integration compared to current-injection-based counterparts[4,5]. However, current NI-EL technologies fall significantly behind mainstream commercially dominant DC-driven light-emitting diodes (DC-LEDs), particularly due to their characteristically limited luminous output and high operating voltages[6].

The first universal challenge faced by all NI-EL configurations is their inherent limitation in generating charges to excite luminescent centers. At the core of a high-field EL device, a light-emitting layer is sandwiched between a duo of dielectric barriers, wherein charge supply relies primarily on tunneling or Schottky emission from the interface traps[4]. The shortcoming in excitation efficiency ($\eta_{\mathrm{exc}}$), typically less than 20%[7-11], persists in the subsequent explorations despite the attempts to adopt metallic particles and doped charge-transporting layers as alternative sources of charges, which shifted the excitation mechanism and modestly increased EL intensity. So far, the pinnacle of NI-EL performance culminated in a luminance of only a few thousand nits, albeit at operating voltages reaching 60 $V_{rms}$[12,13].

Concurrently, NI-EL is generally constrained by the scarcity of suitable light-emitting materials. Early high-field devices relied on impurity-activated ZnS—materials considered substandard for contemporary applications[5]. In the past two decades, exploratory efforts to incorporate highly luminescent excitonic semiconductors with color saturation, including organic molecules, quantum dots (QDs), and low-dimension perovskites[14-16], have yielded limited success in improving NI-EL performance. Fundamentally, the conflict between the field-driven device operation and the excitonic nature of emitters imposes a critical challenge: radiative recombination efficiency ($\eta_{\mathrm{rad}}$) is severely reduced when excitons are under operational electric-field strengths, typically exceeding 1 MV/cm[17,18].

Herein, we present a novel type of NI-EL devices that simultaneously address the two critical challenges. Firstly, we transformed the internal charge-generation mechanism by introducing a pair of state-abundant charge reservoirs, each spatially separated from the emitter, enhancing charge generation by a factor of exceeding 20 and achieving near-unity $\eta_{\mathrm{exc}}$. Secondly, we developed nonmonotonically graded shells for light-emitting QDs, which reduce field-induced exciton quenching and yield high-field $\eta_{\mathrm{rad}}$ of over 90%. These conceptual advancements culminated in RGB NI-EL with bright and efficient emissions. Paradigm-breaking performance metrics include low turn-on thresholds (e.g. 3.7 $V_{rms}$ for red), high luminance intensity with high power efficiency (e.g. 291,628 nits and 302.6 lm/W for green), and superior QD stability compared to QD-LEDs (blue). Minimal dielectric loss, sub-100 ns temporal response, and unprecedentedly high external EL efficiencies (37.1% for red, 45.4% for green, 42.6% for blue) further elucidate the proposed operational mechanism. The results convincingly overcome the inherent bottlenecks of NI-EL, achieving levels competitive with—and often superior to—those of injection-type EL devices across multiple key parameters.

## Results and discussion

### *Device architecture and light-emitting materials*

Figure 1A shows that state-of-the-art NI-EL devices often utilize bulk-doped hole- and electron-transporting layers (HTL and ETL) or high-work-function interlayers to enhance charge generation, yet $\eta_{\mathrm{exc}}$ is ultimately constrained by the density of introduced acceptor or donor states[19,20]. As shown by the schematic diagram in Figure 1B and the cross-sectional transmission electron microscopy image in Figure 1C, the proposed device architecture departs from conventional paradigms by featuring a pair of state-abundant charge reservoirs—n-doped $SnO_2$ nanocrystals (n-type) and p-doped poly-TPD (p-type), respectively. Each reservoir couples to an undoped HTL or ETL via a high-work-function PMA interface layer, forming a tunneling heterojunction that isolates it from the light emitter. A $HfO_2$ barrier between each charge reservoir and its electrode prevents external charge injection. During a positive-bias half-cycle, mobile holes are directly generated within the HTL and transferred to the emitter. This initiates as valence electrons from HTL tunnel across the heterojunction, facilitated by the reduced energy barrier owing to PMA's deep-lying LUMO[21]. The tunneled electrons then occupy the n-type reservoir's LUMO states, where they remain until being released to reoccupy the HTL HOMO during the subsequent negative-bias half-cycle. Likewise, field-generated electrons transfer to the emitter after the p-type reservoir's valence electrons tunnel to ZnO-based ETL via the corresponding heterojunction, leaving holes in its HOMO. As subsequent sections elaborate, $\eta_{\mathrm{exc}}$ amplification primarily stems from the abundance of the electronic states in the charge reservoirs.

In parallel, unlike those developed for DC-LEDs[22-24], the present QDs possess nonmonotonically graded shells, whose radial distributions and band-edge energies are shown in Figures 1D-F. With zero-field of 92–96%, they are expected to stabilize the electron–hole wavefunction against exciton dissociation and surface-bulk nonradiative coupling, thus mitigating emission quenching without compromising charge transfer. We sandwiched QD films between dielectric barriers and quantified the QDs' PL loss. As depicted in Figures 1G-I, the focus is on PL loss under an RMS electric field strength of 0.6 MV/cm, enough for activating EL, and a higher strength of 1.5 MV/cm, typical for achieving high luminance. Taking the green QDs as an example, the control QDs with the regular shell exhibit $\eta_{\mathrm{rad}}$ of only 66.5% and 10.1% under these two strengths, respectively, aligning with prior DC-field measurement[25]. In contrast, employing nonmonotonically graded shells allows for $\eta_{\mathrm{rad}}$ of 93.6% and 78.1% under the same conditions, respectively (see Fig. 1H). Red and blue QDs follow the same trend as the green QDs. In the high field of 1.5 MV/cm, $\eta_{\mathrm{rad}}$ is increased by factors of 7.5, 7.7, and 8.2 for the RGB QDs, respectively, by the nonmonotonically graded-shell design (Figs. 1G–I). The result was further confirmed by measuring the EL of the conventional high-field architecture, as shown in the insets of Figures 1G–I. Compared with the control samples, the RGB QDs developed herein exhibit over six-fold enhancement in EL intensity at the same operating voltage (30 $V_{\mathrm{rms}}$), consistent with the field-dependent PL results.

## *NI-EL performance*

By combining the proposed device architecture with light-emitting QDs, we fabricated RGB NI-EL devices mainly through solution processing. Here, two controls (labeled structures A and B) for each emission color are presented, in addition to our target (labeled structure C). Both structures A and B utilize the conventional QDs and dual PMA interlayers, while B additionally includes the proposed dual charge reservoirs. Figures 2A to 2C show the time-averaged luminance intensity plotted against the amplitude of the 150-kHz sinusoidal AC voltage. Devices with structure A readily exhibit three times the luminance of the brightest reported QD-based NI-EL[13], owing to the PMA-mediated field screening effect (see Extended Data Fig. 1). Beyond this, incorporating the pair of charge reservoirs in structure B universally boosts luminance by ~40-fold ($V$ = 14.1 $V_{\mathrm{rms}}$) for the RGB devices, while lowering threshold voltages ($L_0$ = 1 nit) by ~37%. Further comparison of devices with structures B and C reveals that adopting QDs with nonmonotonically graded shells results in a further ~2.5-fold augmentation of EL intensity and an extra ~15% reduction in threshold voltages.

In Figure 2A, it is noteworthy that the device with structure C, which emits a saturated red color, exhibits a turn-on threshold ($L_0$ = 1 nit) of only 3.7 $V_{\mathrm{rms}}$. This value is half of the lowest reported value[13]. The maximum luminance intensity reaches 48,006 nits at 30 $V_{\mathrm{rms}}$, 9.6 times brighter than red-emitting NI-EL benchmarks[12,13]. Figure 2D shows that the power efficiency ($\eta_{\mathrm{power}}$) of 28.3 lm/W further increases to 40.3 lm/W at lower frequencies, a 40.3-fold enhancement over the best reported value (see Fig. 3A)[12]. Turning to Figures 2B and 2E, the green-emitting device with structure C also demonstrates ultralow-voltage operation with turn-on at merely 3.9 $V_{\mathrm{rms}}$, delivering 1000 nits luminance at 6.3 $V_{\mathrm{rms}}$. The maximum luminance exceeds 206,000 nits, establishing a 41.4-fold enhancement over any reported NI-EL devices[12]. More intriguingly, the already-remarkable $\eta_{\mathrm{power}}$ of 162.2 lm/W (150 kHz) further increases to 214.6 lm/W at the frequency of 30 kHz. This figure is 21.5 times greater than that of any reported NI-EL (see Fig. 3A)[10].

Figures 2C and 2F showcase the first-reported NI-EL with saturated blue emission. Unlike previous attempts that yielded very weak emissions, e.g. 1 nit at the cost of 100 $V_{\mathrm{rms}}$[26], the device with structure C combines a turn-on threshold of only 5.3 $V_{\mathrm{rms}}$, a luminance of up to 51,435 nits, and a $\eta_{\mathrm{power}}$ of up to 32.3 lm/W. Moreover, this device effectively mitigates the instability issue of blue QDs, a critical bottleneck for DC-QLED. Figure 2G compares the PL degradation measured from the QD layer within two types of EL devices. After the initial decrease due to recoverable charging in QDs, the PL intensity exhibits a 40% degradation during the testing period, largely attributed to QD degradation linked to deep traps within them[27]. In contrast, the QDs in the NI-EL device exhibit a degradation of only 15.6% while retaining the same initial EL intensity (10,000 nits). We attribute this advantage to the AC operational mode, which promotes charge detrapping to slow QD degradation while ensuring electrode stability by eliminating external charge injection.

These devices with structure C retain near-Lambertian emission characteristics, suggesting that the contribution of light-extraction gain to bright and efficient emission is likely negligible. Upon integration of a hemispherical lens to improve light outcoupling, the RGB devices (denoted as structure C + lens) exhibit further luminance enhancement factors of 1.50, 1.41, and 1.45[28], reaching maximum values of 72,009, 291,628, and 74,581 nits, and achieving power efficiencies of 60.4, 302.6, and 46.9 lm/W, respectively.

We further assessed the significance of these results by reviewing diverse EL device benchmarks. First, Figure 3A shows that our devices hold a clear and substantial advantage over existing NI-EL benchmarks across all key metrics—turn-on threshold, luminance, and power efficiency—regardless of configuration or emitter material, due to their unmatched $\eta_{\mathrm{exc}}$ and $\eta_{\mathrm{rad}}$ [8,10,12,13,19,29-34]. Beyond this, the extent of performance breakthrough warrants a re-evaluation of the supposed inherent performance gap between this technology and injection-type EL. As shown in Figure 3B, our NI-EL demonstrates an appreciable superiority in luminance intensity (2.8-fold for red, 11.3-fold for green, and 24.9-fold for blue) and power efficiency (3.9-fold for red, 4.2-fold for green, and 15.1-fold for blue) over best-reported metrics for injection-type AC-driven EL devices[35-41]. This indicates that the magnitude and efficiency of the internal charge generation likely surpass that can be supplied externally through one or both Ohmic-contact electrodes, while generating much less dielectric loss, as discussed below. More encouragingly, the NI-EL devices presented herein can effectively compete with injection-type DC-LEDs based on QDs and organic emitters in the aspects of operating voltage and luminance intensity (see Fig. 3C and Fig. 3D) [42-52]. The green device, in particular, achieves a power efficiency (302.6 lm/W) surpassing reported QD- and organic-based DC-LEDs.

## *NI-EL mechanism*

The above results indicate that the novel device architecture not only contributes substantially to the overall performance gain but may also signal a shift in the excitation mechanism. Here, we conducted a comparative analysis to elucidate the functions of the charge reservoirs. The discussions on tunneling junctions and undoped HTL/ETL are provided in the Extended Data Fig. 1 and Extended Data Fig. 2.

Figures 4A-C illustrate the temporal voltage bias, *v(t)*, current density, *j(t)*, and EL intensity, *e(t)*, with all samples labeled as in Figures 2A-C. In structure A, the interfacial PMA provides a limited number of acceptor states to assist electron tunneling. In comparison, introducing dual charge reservoirs in structure B substantially increases the density of accessible states, doubling the magnitude of *i(t)*—a result not markedly affected by further QD material upgrades (structure C). Nevertheless, as the *i(t)* magnitude alone does not sufficiently reflect changes in charge generation capability, analysis of the temporal relationship between *j(t)* and *v(t)* remains essential. As seen from the middle panels, the RGB devices universally show increased *v(t)*-*j(t)* phase angles upon the addition of dual charge reservoirs (structures B and C). In particular, the red and green devices exhibit *j(t)*-*v(t)* phase angles of >89.3 and 88.4 degrees at 10 kHz and 100 kHz, respectively (Fig. 4D). This contrasts sharply with previously reported AC-driven configurations, which show much lower phase angles regardless of charge injection[12,53]. The evidently capacitive response highlights that the dual charge reservoirs markedly reduce relaxation losses compared to conventional interface trapping/release mechanisms.

As shown in the lower panels of Figures 4A-C, the dual charge reservoirs enable a universal 40-fold enhancement in *e(t)* amplitudes, accompanied by improved synchronization to *v(t)*. Closer examination further reveals that these large-area devices (4 $mm^2$) have a response time, defined as the *e(t)-v(t)* peak delay, of below 100 nanoseconds (see response to square wave in Extended Data Fig. 3). This sharply contrasts with those spanning from a few microseconds to milliseconds observed in AC- or DC-driven benchmarks across configurations[37,54]. The markedly accelerated responses of *i(t)* and *e(t)* collectively demonstrate an unprecedented level of field-driven excitation, where charges are created in direct contact with the QDs, with their generation magnitudes and rates amplified by the charge reservoirs' abundant electronic states. By additionally adopting QDs with robust high-field fluorescence, accumulative enhancements in *e(t)* reach 70- to 120-fold without reducing the *e(t)-v(t)* synchronization (see Figs. 4A-C).

The frequency-dependent EL characteristics of structure C further corroborate the remarkably efficient excitation process. Figures 2D-F and Extended Data Fig. 4 show that our NI-EL devices exhibit decent luminance and efficiency at the maximum frequency of 150 kHz, limited by the power source. As frequency decreases, luminance declines with the duty cycle while electrical efficiency rises with the *j(t)-v(t)* phase angle, together producing a peak $\eta_{\mathrm{power}}$ near 30 kHz. We converted $\eta_{\mathrm{power}}$ into external efficiency of each EL pulse ($\eta_{\mathrm{EL}}$, defined as radiant energy divided by electrical energy consumed during each cycle). As shown in Fig. 4e, the $\eta_{\mathrm{EL}}$ peaks at 24.8% for red, 32.4% for green, and 29.6% for blue; with an attached hemispherical lens, these values further increase to 37.1%, 45.4%, and 42.6%, respectively. Although such a metric has rarely been reported in prior work, its significance for NI-EL mirrors that of external quantum efficiency for injection-type EL. To our best knowledge, the $\eta_{\mathrm{EL}}$ results reported here, which approach the light-outcoupling limit, exceed all previous NI-EL records by a wide margin.

Further incorporating the measured $\eta_{\mathrm{rad}}$ (Figs. 1G-I), the calculated light outcoupling efficiency ($\eta_{\mathrm{out}}$, see Extended Data Fig. 5), and the relation:

$$\eta_{\mathrm{EL}} = \eta_{\mathrm{exc}}\eta_{\mathrm{rad}}\eta_{\mathrm{out}} \tag{1}$$

we find that $\eta_{\mathrm{exc}}(f)$ dominates the frequency-dependent EL characteristics, as shown in Figure 4F. The resulting $\eta_{\mathrm{exc}}$ peaks (86.3% for red, 94.0% for green, 92.6% for blue) of structure C uniformly exhibit a greater-than-20-fold enhancement over the corresponding structure A control (see summary across device structures in Extended Data Table 1). Moreover, they significantly surpass those of conventional high-field excitation of ZnS ($\eta_{\mathrm{exc}}$<20%)[7-9,30,31] and relatively recent NI-EL developments ($\eta_{\mathrm{exc}}$<20%)[10-12], while even rivaling those of the most efficient DC-LEDs ($\eta_{\mathrm{exc}}$>90) [27,49,55]. In particular, the near-unity $\eta_{\mathrm{exc}}$ of our green device, combined with its high $\eta_{\mathrm{rad}}$, is key to achieving the record-eclipsing $\eta_{\mathrm{power}}$ and $\eta_{\mathrm{EL}}$.

## Conclusion

To overcome the longstanding limitations of NI-EL, this work introduces advances in both device architecture and luminescent materials. The resulting RGB devices can consistently operate at voltages below half of the established benchmarks, with the red device, in particular, achieving a turn-on threshold of only 3.7 $V_{\mathrm{rms}}$. Their luminance universally exceeds respective benchmark levels by a factor ranging from 9.6 to 103.6, as highlighted by the green device, which outputs a time-averaged luminance of over 291,628 nits and a power efficiency of 302.6 lm/W with outcoupling enhancement, substantially surpassing prior NI-EL records and rivaling the most efficient DC-LEDs. Furthermore, the first QD-based blue NI-EL not only achieves high luminous efficiency (e.g. 46.9 lm/W) but also exhibits QD stability superior to that of its DC-LED counterpart under comparable operating conditions. Mechanistic studies attribute the performance breakthroughs to two key factors. First, charge reservoirs integrated with tunneling heterojunctions achieve a >20-fold enhancement in charge generation and field-driven excitation efficiency above 94%, as evidenced by minimal dielectric loss, sub-100 ns response, and record-eclipsing EL efficiencies (37.1% for red, 45.4% for green, and 42.6% for blue). Second, QDs with nonmonotonically graded shells exhibit an over 7-fold increase in high-field fluorescence, enabling radiative efficiency exceeding 90% under operational conditions. These advances not only break fundamental bottlenecks in NI-EL, but also position it to compete with leading EL technologies in certain application scenarios.

## Figure Legends

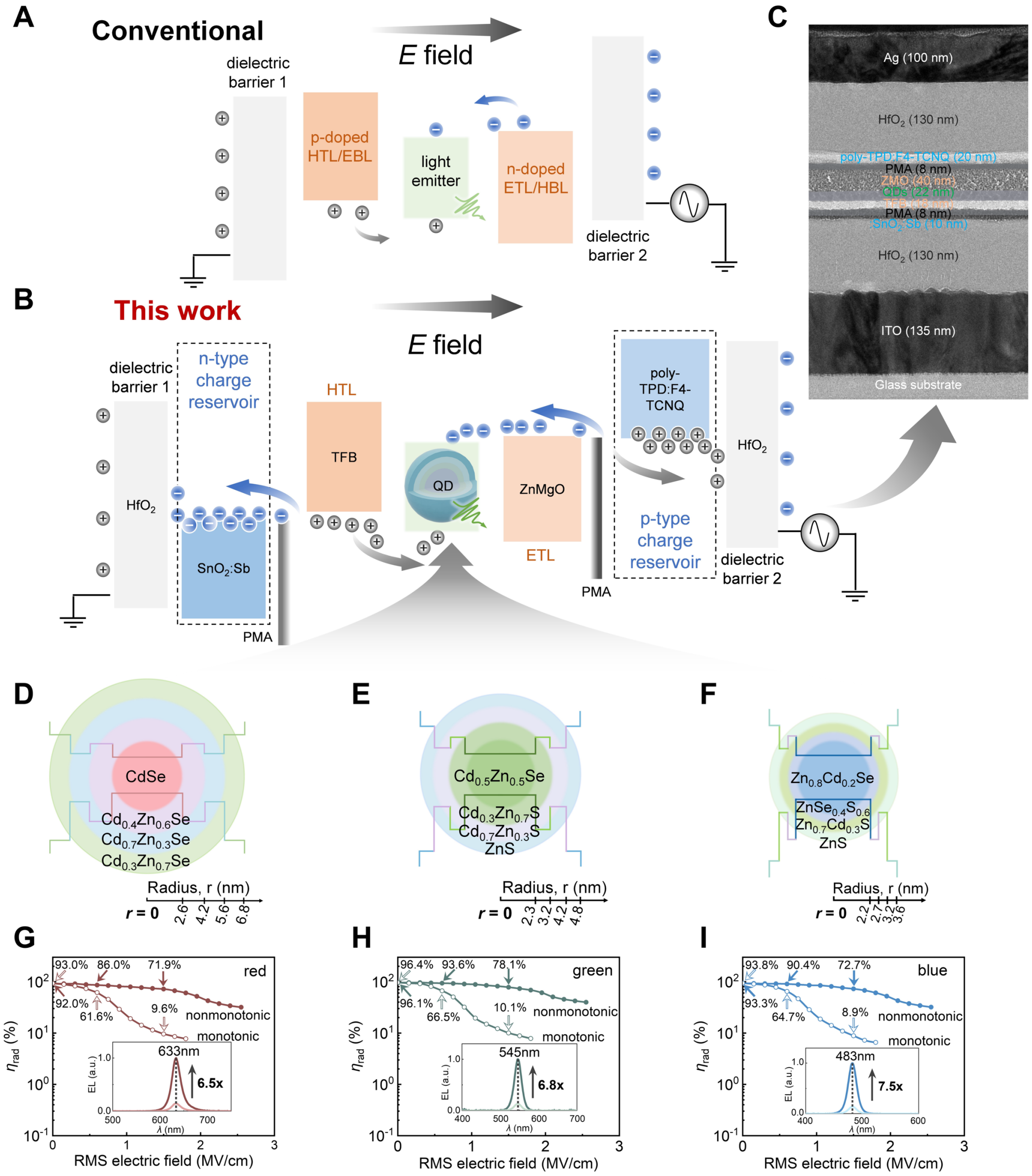

**Fig. 1. Device architecture and emissive materials proposed in this work. (A)** Schematic energy-level diagram of a conventional device architecture with limited excitation efficiency ($\eta_{\text{exc}}$). **(B)** Schematic energy-level diagram and the electroluminescence (EL) mechanism of our proposed device architecture, highlighting the pair of charge reservoirs integrated with tunneling junctions as the key to breaking through the longstanding bottleneck of $\eta_{\text{exc}}$. **(C)** Cross-sectional TEM image of a green device with structure C, showing the material and thickness of each functional layer. **(D-F)** Approximate energy band diagrams and radial compositions of the red (D), green (E), and blue (F) quantum dots (QDs) with the proposed nonmonotonically graded shells to suppress field-induced emission quenching. **(G-I)** Radiative recombination efficiency ($\eta_{\text{rad}}$) of the red (G), green (H), and blue (I) colloidal QDs, respectively. Calibrated zero-field internal $\eta_{\text{rad}}$ is measured from solid films using a standard method that corrects for optical losses. Samples with conventional high-field structure (electrode/barrier/QD/barrier/electrode) are used for measuring field-dependent relative $\eta_{\text{rad}}$. The Insets show the EL spectra measured from the same samples.

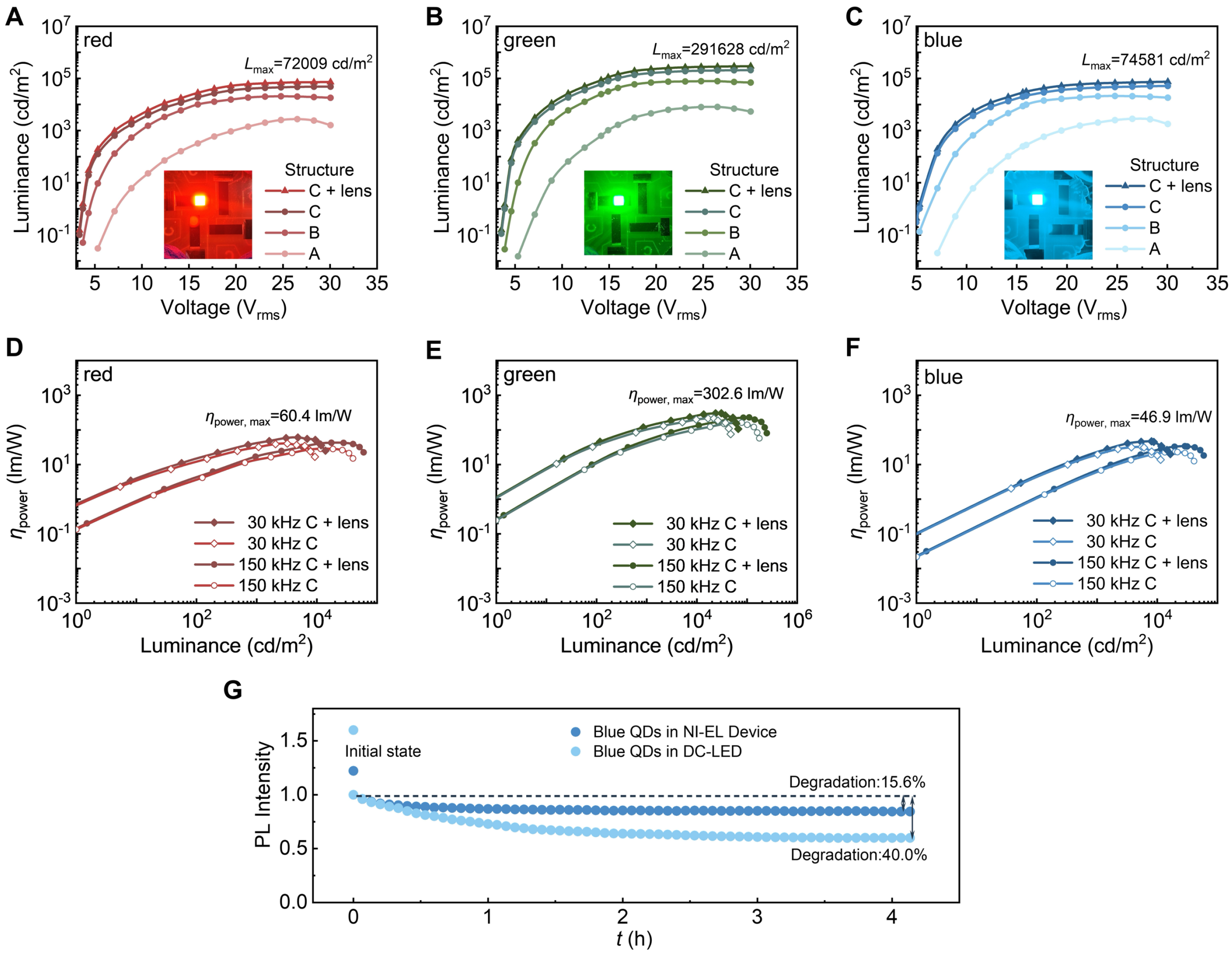

**Fig. 2. Non-injecting electroluminescence (NI-EL) device performance achieved in this work. (A-C)** Luminance-voltage ($L$-$V$) characteristics of the red (A), green (B), and blue (C) NI-EL devices, respectively, measured under AC bias at a frequency of 150 kHz. The insets show photographs of the corresponding device emission in the operating. **(D-F)** Power efficiency ($\eta_{\text{power}}$) as a function of luminance for the red (D), green (E), and blue (F) NI-EL devices, respectively, driven by AC bias at frequencies of 30 kHz and 150 kHz. **(G)** Photoluminescence (PL) decay of the blue quantum-dot (QD) layer in the operating devices, both with the same initial luminance of 10,000 nits. Here, all the devices with structures A and B comprise the first dielectric barriers, the first PMA layer, the hole-transporting layer, the emitter, the electron-transporting layer, the second PMA layer, and the second dielectric barrier. Structures A and B employ monotonically graded QDs as the emissive layer, whereas structure C adopts nonmonotonically graded QDs. Structures B and C additionally feature a pair of charge reservoirs. Structure C + lens represents structure C with an attached hemispherical lens for enhanced light extraction.

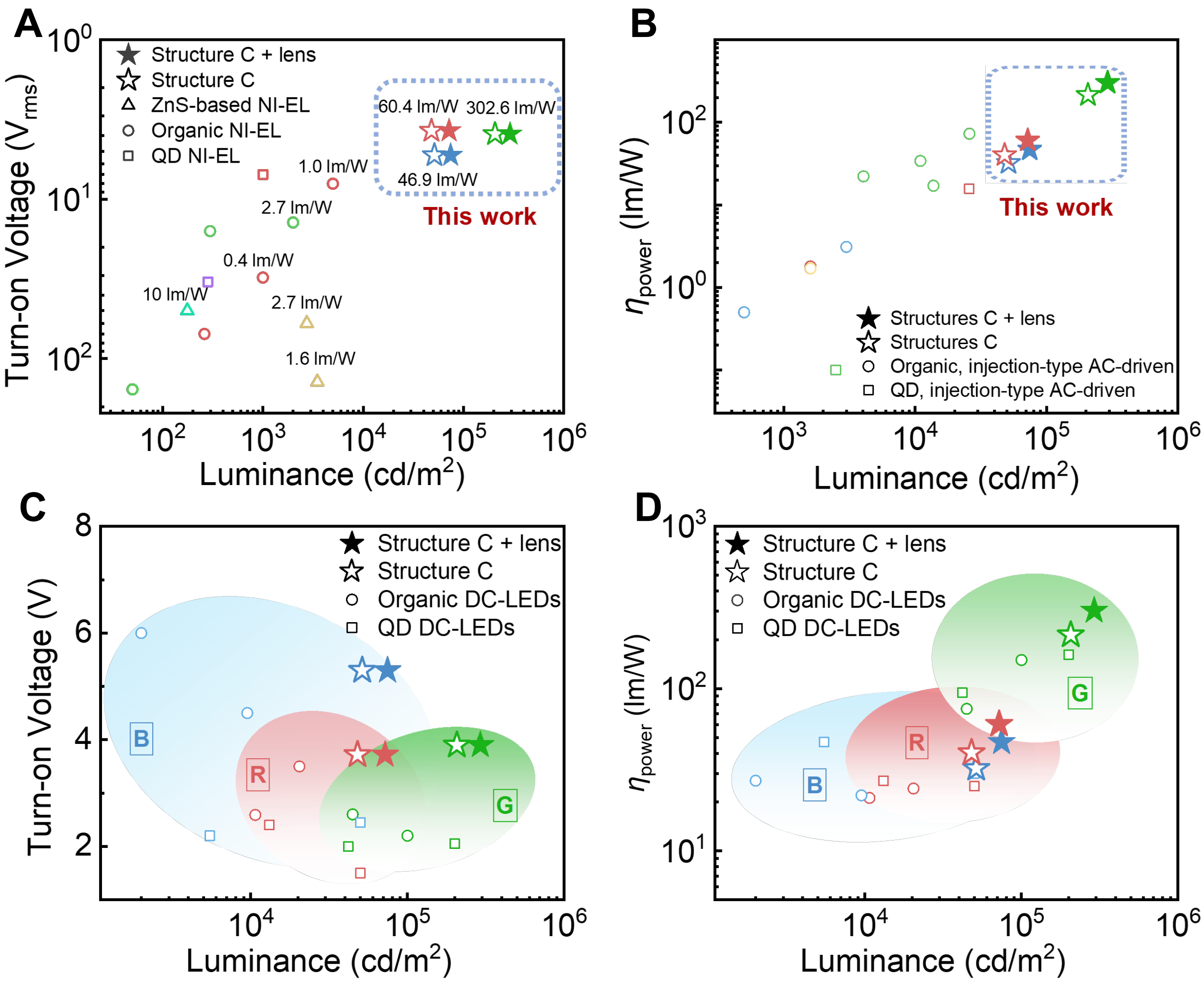

**Fig. 3. Performance comparison across electroluminescent device categories. (A)** Comparison of maximum luminance and turn-on voltage with previously reported NI-EL devices. The physical quantity in lumens per watt adjacent to each symbol represents the power efficiency of the corresponding device ($\eta_{\text{power}}$). **(B)** Comparison of maximum luminance and maximum power efficiency with previously reported injection-type AC-driven devices. **(C)** Comparison of turn-on voltage with representative DC-driven light-emitting diodes (DC-LEDs). The turn-on voltage is defined as the root-mean-square voltage ($V_{\text{rms}}$) in AC-driven devices and as the nominal DC turn-on voltage in DC-LEDs. **(D)** Comparison of maximum power efficiency and maximum luminance with previously reported DC-LEDs. In the above panels, the open (filled) markers denote devices without (with) a hemispherical lens. The stars represent devices reported in this work. The triangles, circles, and squares represent previously reported devices based on ZnS-based materials, organic emitters, and quantum dots. Symbol colors correspond to the emission colors of their respective devices.

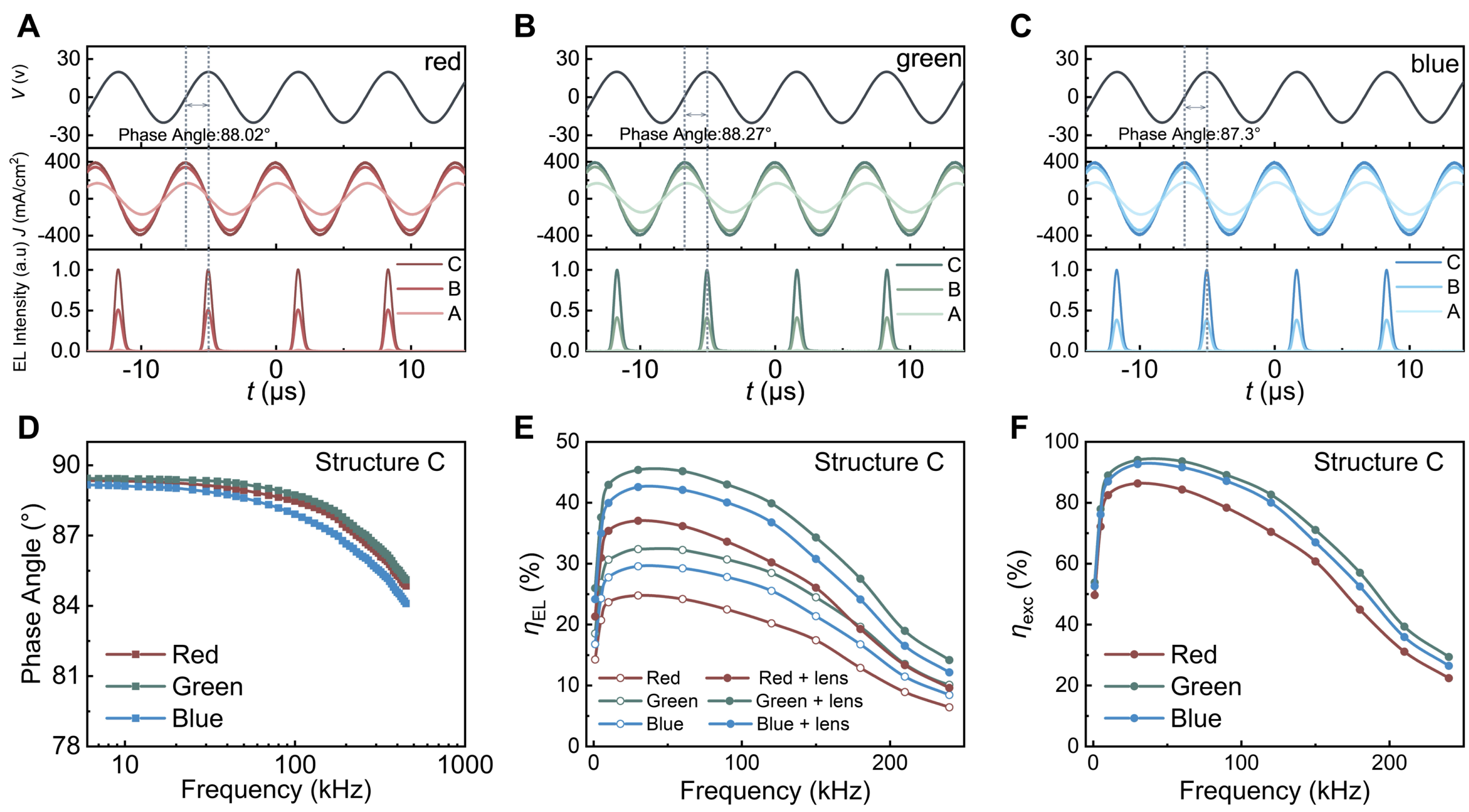

**Fig. 4. Temporal and frequency-dependent electroluminescence (EL) characteristics. (A-C)** Temporal profiles of voltage bias *v(t)*, current density *j(t)*, and EL intensity *e(t)* for the red (A), green (B), and blue (C) NI-EL devices, respectively. **(D)** Frequency-dependent *j(t)-v(t)* phase angles. The measurements were conducted under a bias of 14.1 $V_{rms}$. **(E)** Frequency-dependent external efficiency of each EL pulse ($\eta_{\mathrm{EL}}$). **(F)** Frequency-dependent excitation efficiency ($\eta_{\mathrm{exc}}$). In the measurements of electroluminescence efficiency, the voltage was set to the value that maximized the efficiency of each device.

## Extended data figure/table legends

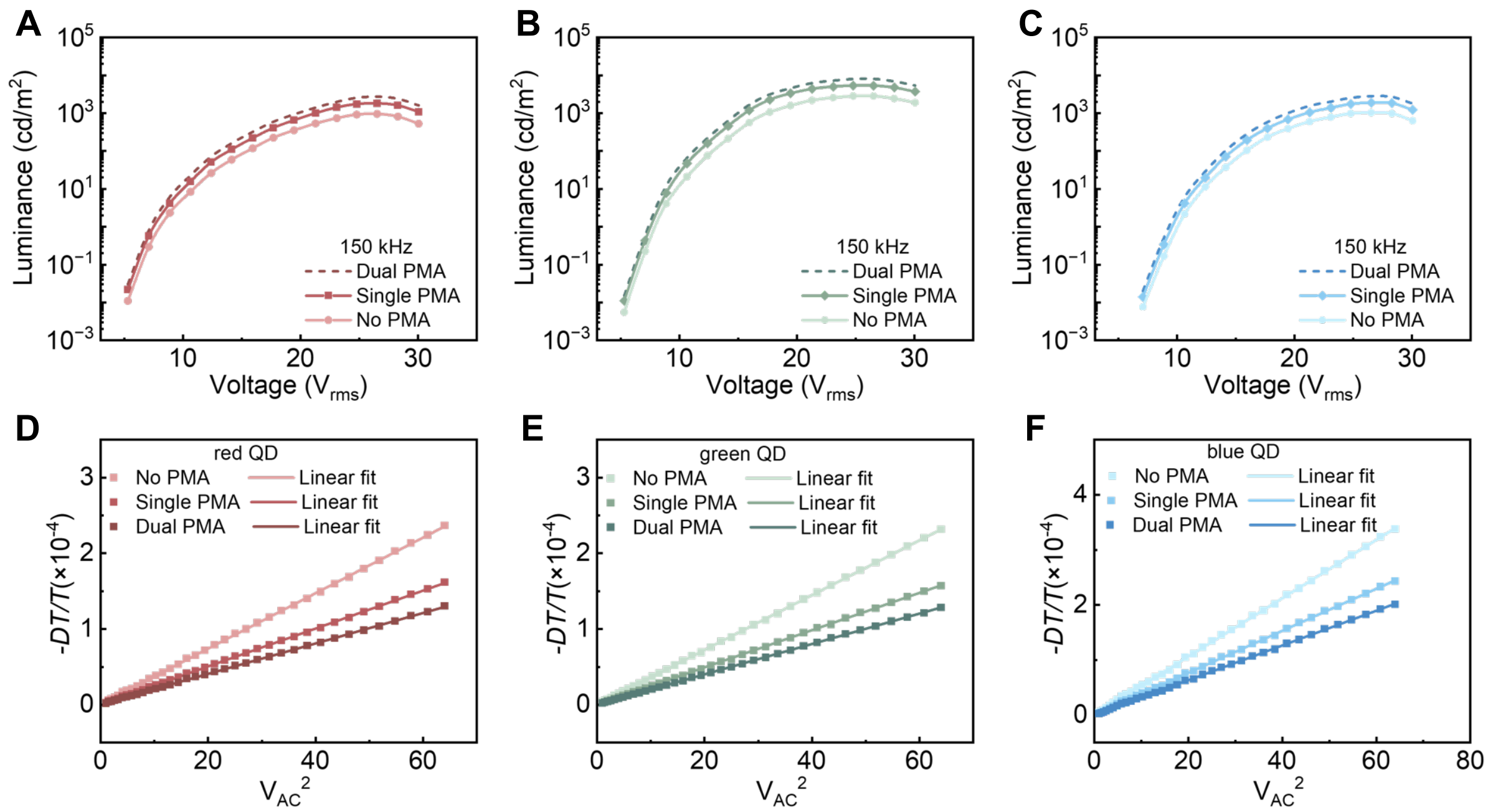

**Extended Data Fig. 1 | Field-screening effect of the PMA interlayer. (A-C)** Luminance-voltage (*L-V*) curves of the red, green, and blue devices, respectively; **(D-F)** corresponding electroabsorption (EA) spectra. In each panel, the three curves represent devices without PMA, with a single PMA interlayer, and with dual PMA interlayers, respectively. The single-PMA and dual-PMA devices each contain a single tunneling heterojunction, whereas the PMA-free device contains no tunneling heterojunction. The PMA-free device adopts a device architecture representative of the brightest reported organic-based NI-EL device and the brightest reported QD-based NI-EL device and is therefore used as the benchmark, whereas the single-PMA device serves as a reference device for investigating the PMA-mediated field-screening effect. Dual PMA interlayers are expected to reduce field-induced emission quenching, owing to their ability to form highly depleted regions on both HTL and ETL surfaces. Quantum-confined Stark effect measurements confirm this mechanism, showing a linear scaling of the second-harmonic electroabsorption signal with the square of the AC field amplitude in the QD layer. The dual-PMA and single-PMA configurations reduce the electric field across the QDs by approximately 45% and 32%, respectively. Correspondingly, their luminance values are approximately 2.8 and 1.9 times that of the PMA-free benchmark device. This enhanced field-screening effect primarily accounts for the superior performance of structure A, which achieves nearly three times the luminance of the brightest previously reported QD-based NI-EL device[13].

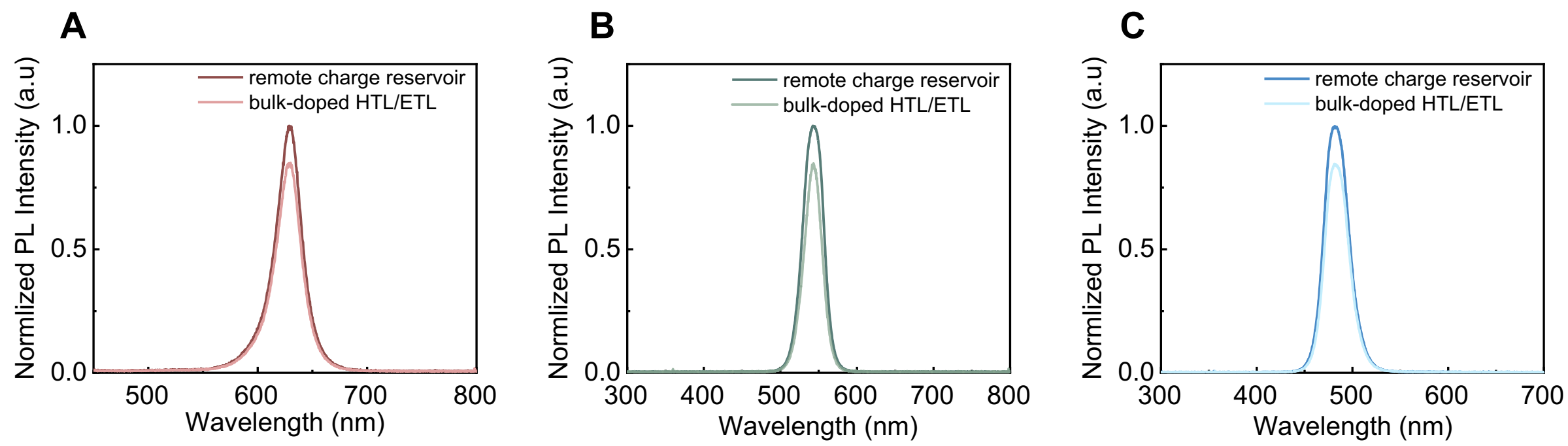

**Extended Data Fig. 2 | Comparison between bulk-doped HTL/ETL and spatially remote charge-reservoir configurations. (A-C)** QD fluorescence intensity of the red, green, and blue devices, respectively. In each panel, the two curves correspond to the mildly doped HTL/ETL structure and the spatially remote charge-reservoir structure. For all three emission colors, the remote charge-reservoir structures exhibit noticeably higher PL intensities compared to their bulk-doped counterparts. This confirms that spatially separating the dopants from the emissive layer effectively suppresses the exciton quenching typically induced by direct contact with the doped transport layers.

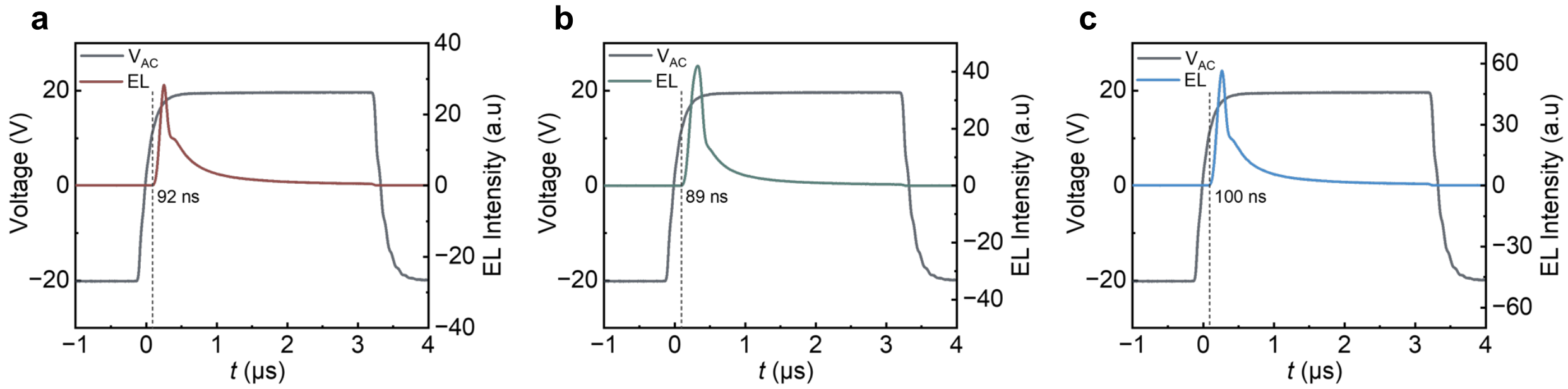

**Extended Data Fig. 3 | Transient EL responses of NI-EL devices. a-c,** red (a), green (b), and blue (c) devices, driven by a square wave at 150 kHz and 14.1 $V_{rms}$. In each panel, the gray trace (corresponding to the left y-axis) represents the applied driving voltage, and the colored trace (corresponding to the right y-axis) depicts the transient EL intensity. The zero-time point ($t = 0$) is defined as the zero-crossing point of $v(t)$. The vertical dashed line and the labeled values (92, 89, and 100 ns for the red, green, and blue devices, respectively) indicate the EL emission response delay time relative to the driving voltage rise edge ($t = 0$).

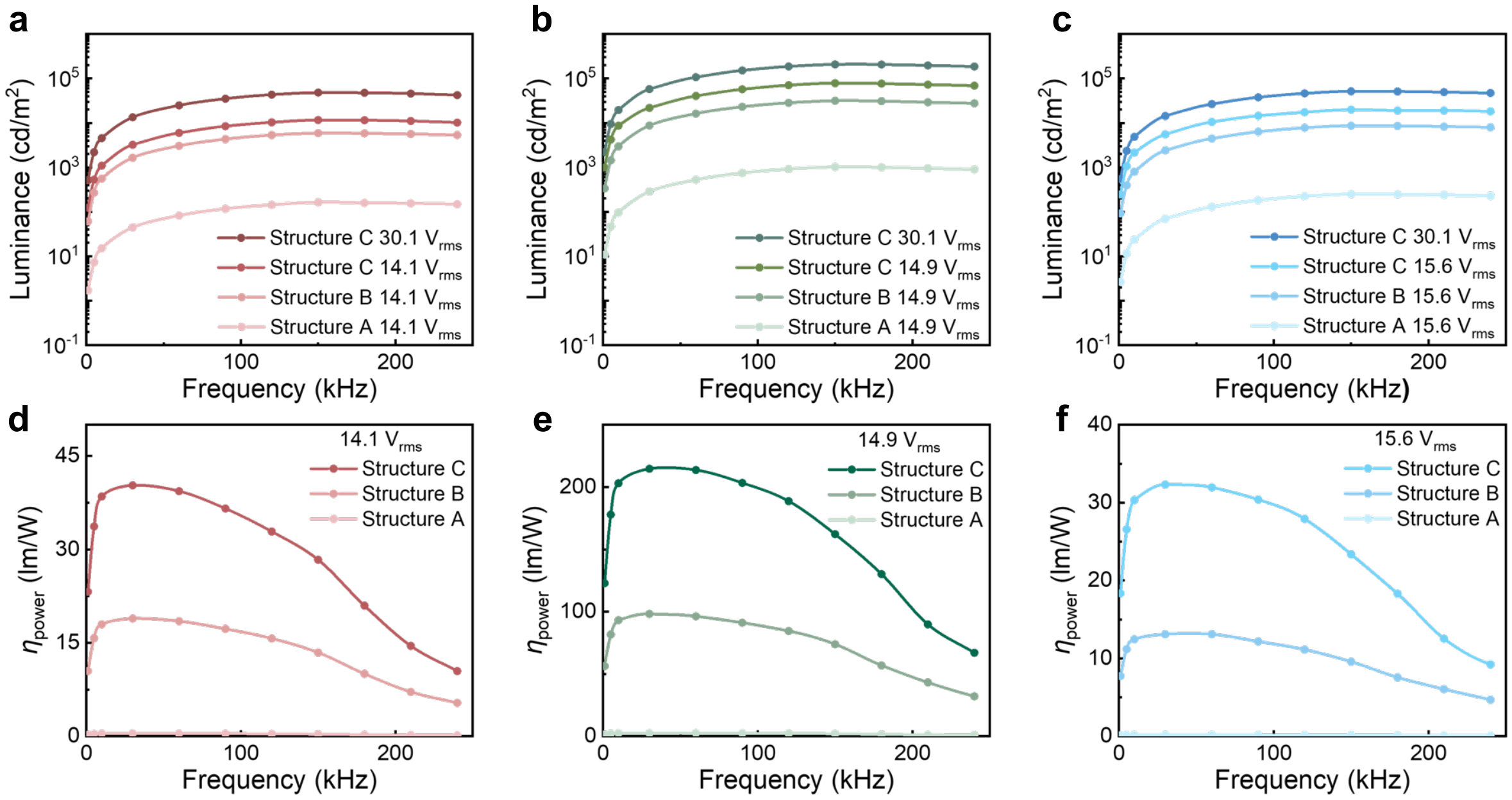

**Extended Data Fig. 4 | Frequency-dependent luminance and power efficiency ($\boldsymbol{\eta}_{\mathbf{power}}$) characteristics of the NI-EL devices. a-c,** Luminance as a function of driving frequency for the red (a), green (b), and blue (c) devices with structures A, B, and C, measured at their respective power-efficiency-optimized driving voltages, and additionally shows the highest-brightness performance of the optimized structure C at a higher voltage (30.1 $V_{rms}$). **d-f, $\boldsymbol{\eta}_{\mathbf{power}}$** versus driving frequency for the corresponding red (d), green (e), and blue (f) devices with structures A, B, and C, evaluated under the efficiency-optimized voltage condition for each structure. In each panel, results from structures A, B, and C are shown for comparison.

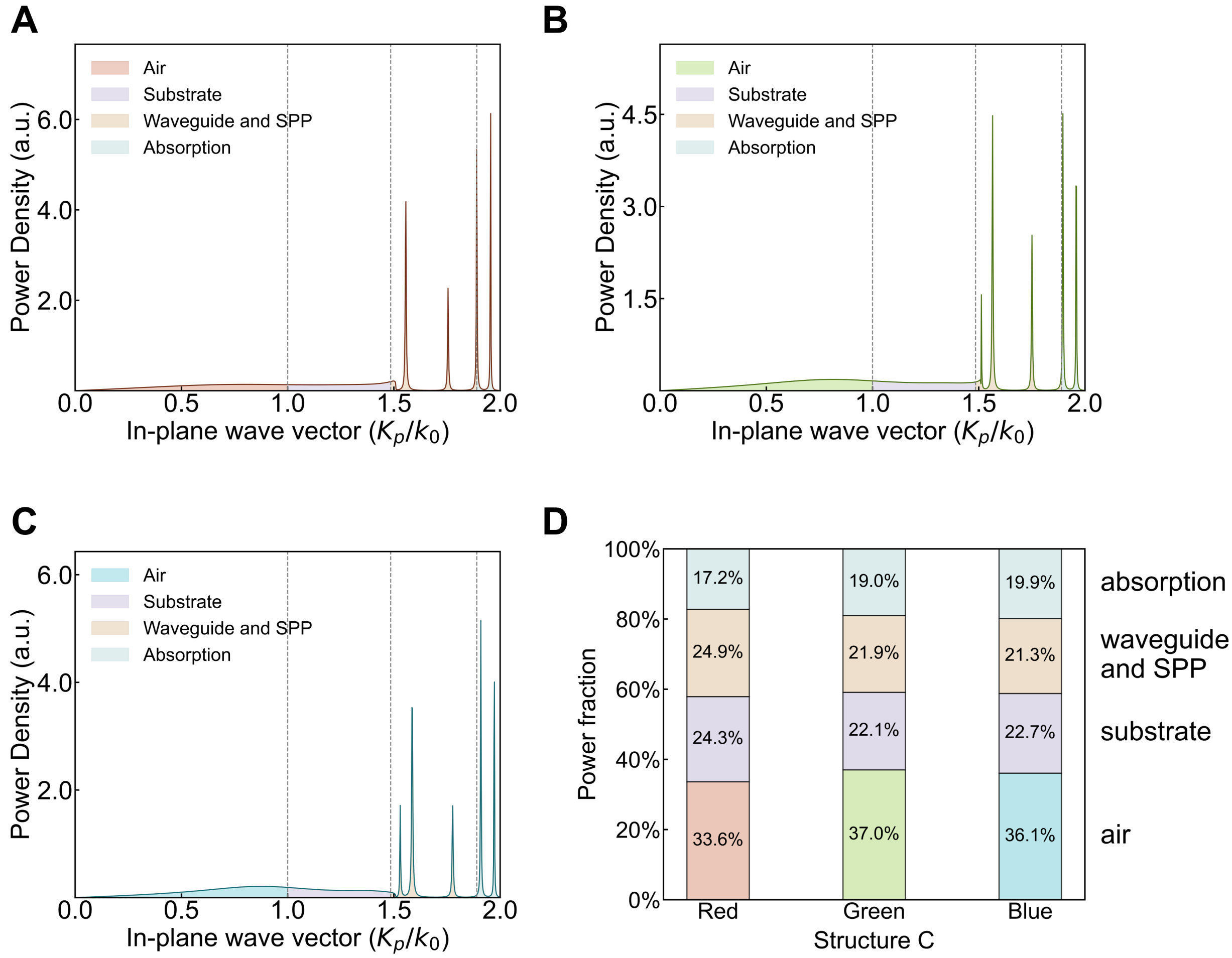

**Extended Data Fig. 5 | Optical modeling of NI-EL devices. (A-C)** Power-density distributions versus the in-plane wavevector ($k_p/k_0$) for red, green, and blue devices. Here, $k_p$ is the in-plane wavevector and $k_0$ is the wavevector in vacuum. The vertical dashed lines demarcate the boundaries for integrating different optical modes. **(D)** Calculated power fractions of each optical channel obtained by integrating the corresponding regions in a–c. Here, "air", "substrate", "waveguide and SPP", and "absorption" denote the fractions of power out-coupled to air (that is, the light-extraction efficiency), trapped in the glass substrate (recoverable with a hemispherical lens), absorbed by surrounding media, and coupled to waveguide modes or surface plasmon polaritons (SPPs), respectively.

**Extended Data Table 1 | Summary of NI-EL performance across all structures demonstrated in this work.**

| Red NI-EL devices in this work | | | | | | | |
|---|---|---|---|---|---|---|---|
| **Device structure** | **Turn-on voltage ($V_{rms}$)** | **Maximum luminance ($cd/m^2$)** | $\eta_{power}$ **(lm/W)** | $\eta_{EL}$ **(%)** | $\eta_{exc}$ **(%)** | $\eta_{rad}$ **(%)** | $\eta_{out}$ **(%)** |
| Structure A | 7.2 | 2,730 | 0.8 | 0.5 | 4.2 | 36.1 | 31.8 |
| Structure B | 4.4 | 20,526 | 21.7 | 13.3 | 81.2 | 48.9 | 33.6 |
| **Structure C** | **3.7** | **48,006** | **40.3** | **24.8** | **86.3** | **85.4** | **33.6** |
| **Structure C + lens** | **3.7** | **72,009** | **60.4** | **37.1** | **86.3** | **85.4** | **50.4** |

| Green NI-EL devices in this work | | | | | | | |
|---|---|---|---|---|---|---|---|
| **Device structure** | **Turn-on voltage ($V_{rms}$)** | **Maximum luminance ($cd/m^2$)** | $\eta_{power}$ **(lm/W)** | $\eta_{EL}$ **(%)** | $\eta_{exc}$ **(%)** | $\eta_{rad}$ **(%)** | $\eta_{out}$ **(%)** |
| Structure A | 7.4 | 8,037 | 4.0 | 0.6 | 4.7 | 35.3 | 35.7 |
| Structure B | 4.7 | 78,098 | 107.2 | 16.2 | 91.2 | 48.0 | 37.0 |
| **Structure C** | **3.9** | **206,829** | **214.6** | **32.4** | **94.0** | **93.1** | **37.0** |
| **Structure C + lens** | **3.9** | **291,628** | **302.6** | **45.4** | **94.0** | **93.1** | **52.2** |

| Blue NI-EL devices in this work | | | | | | | |
|---|---|---|---|---|---|---|---|
| **Device structure** | **Turn-on voltage ($V_{rms}$)** | **Maximum luminance ($cd/m^2$)** | $\eta_{power}$ **(lm/W)** | $\eta_{EL}$ **(%)** | $\eta_{exc}$ **(%)** | $\eta_{rad}$ **(%)** | $\eta_{out}$ **(%)** |
| Structure A | 9.3 | 2,816 | 0.5 | 0.4 | 4.6 | 25.8 | 34.9 |
| Structure B | 6.2 | 21,208 | 14.9 | 13.6 | 88.8 | 42.6 | 36.1 |
| **Structure C** | **5.3** | **51,435** | **32.3** | **29.5** | **92.6** | **88.4** | **36.1** |
| **Structure C + lens** | **5.3** | **74,581** | **46.9** | **42.6** | **92.6** | **88.4** | **51.8** |

Note:
Unless otherwise noted, these efficiency-related parameters were evaluated at the bias voltage that maximized the electroluminescence efficiency of each device.